\documentclass[british,english]{revtex4-1}
\usepackage[T1]{fontenc}
\usepackage[latin9]{inputenc}
\usepackage{units}
\usepackage{mathtools}
\usepackage{amsmath}
\usepackage{amssymb}
\usepackage{stackrel}
\usepackage{esint}

\makeatletter
\usepackage[dvipsnames]{xcolor}
\usepackage[unicode=true,
 bookmarks=false,
 breaklinks=false,pdfborder={0 0 1},%backref=section,
 colorlinks=false]
 {hyperref}
\hypersetup{
 colorlinks,citecolor=cyan,linkcolor=blue,urlcolor=cyan}
\usepackage{breakurl}

\makeatother

\usepackage{babel}

\begin{document}

\title{Non-Equilibrium Steady State of the Lieb-Liniger model: \\
exact treatment of the Tonks Girardeau limit}

\author{Spyros Sotiriadis}

\address{Department of Physics, Faculty of Mathematics and Physics, University
of Ljubljana, Ljubljana, Slovenia}

\begin{abstract}
Aiming at studying the emergence of Non-Equilibrium Steady States
(NESS) in quantum integrable models by means of an exact analytical
method, we focus on the Tonks-Girardeau or hard-core boson limit of
the Lieb-Liniger model. We consider the abrupt expansion of a gas
from one half to the entire confining box, a prototypical case of
inhomogeneous quench, also known as ``geometric quench''. Based
on the exact calculation of quench overlaps, we develop an analytical
method for the derivation of the NESS by rigorously treating the thermodynamic
and large time and distance limit. Our method is based on complex
analysis tools for the derivation of the asymptotics of the many-body
wavefunction, does not make essential use of the effectively non-interacting
character of the hard-core boson gas and is sufficiently robust for
generalisation to the genuinely interacting case.
\end{abstract}

\maketitle

\tableofcontents

\section{Introduction}

The derivation of macroscopic physical laws that govern non-equilibrium
phenomena from the underlying microscopic particle interactions described
by quantum physics is one of the fundamental open questions of statistical
and mathematical physics. In this quest the problem of quantum transport
between two halves of a system, initially at equilibrium with different
thermodynamic parameters and then joined together, is of central interest.
Mathematical physics studies \citep{math:Tasaki2001,math:Araki-Ho,Ruelle,math:Jaki2002,math:Ogata2002,math:Aschbacher2003,Spohn_1977}
followed by an increasing number of theoretical physics works \citep{BernardDoyon,DeLuca2013,Collura2014,CFT:Bernard-Doyon_2014,rev:Bernard2016,Viti2016,Kormos2017,diff:Stephan2017,Moosavi_19}
supported by the development of powerful numerical techniques for
the simulation of quantum dynamics \citep{num:1,num:2,num:3,num:4,num:6,num:7,num:diff1,diff:Misguich2017}
and experimental advances in the observation of quantum many-body
dynamics \citep{exp:1,exp:2,exp:3,exp:4} have kept the attention
on this problem alive and growing for decades.

In more general terms this physical protocol, more recently known
also as ``inhomogeneous quantum quench'', refers to the dynamics
of an extended closed quantum system that is initially prepared in
a macroscopically inhomogeneous state and let to evolve under a homogeneous
Hamiltonian. A typical case of an inhomogeneous initial state is the
one corresponding to the partitioning protocol, outlined above, where
a system is initially split in two disconnected halves at thermal
equilibrium and different temperature or particle density \citep{BernardDoyon,CFT:Bernard-Doyon_2014,rev:Bernard2016}.
Other suitable choices include equilibrium (ground or thermal) states
with a ``local temperature'' (or particle or energy density) profile
varying from one value in the left half to a different one in the
right, in either a smooth or a sharp way \citep{CFT:Sotiriadis,CFT:Moosavi,CFT:Moosavi2},
or quantum gas expansion protocols, also called geometric quenches,
in which a system of atoms expands from a smaller to a larger box
\citep{Mossel-Caux}. 

A first understanding of quantum transport following an inhomogeneous
quench can be drawn from the study of non-interacting systems \citep{Ruelle,DeLuca2013,Collura2014,Viti2016,Lancaster2016,Kormos2017,Dubail2017,Moosavi_20}.
For such systems it has been shown that a \emph{Non-Equilibrium Steady
State} (NESS) emerges at large times, which provides a complete description
of the asymptotic values of local observables, like the particle density
and current. However, while results for non-interacting systems have
been rigorously derived and known for years, the study of the effects
of interactions, which are obviously expected to play a significant
role in transport, turned out to be a hard problem and progress was
slow. A breakthrough came recently from the study of integrable interacting
systems \citep{GHD_Doyon,GHD_Fagotti-Bertini}, which are characterised
by the presence of an infinite number of local conservation laws.
It was postulated that a \emph{Generalised Hydrodynamics} (GHD) theory,
based upon an infinite set of continuity equations associated to the
conservation laws, provides an exact description of the NESS in such
systems. The predictions of this conjecture have been tested numerically
with great success both in spin chains and in quantum gas models solvable
by Bethe Ansatz \citep{GHD_refs:Piroli,GHD_refs:Collura,GHD_refs:Doyon-Caux-Dubail-Konik,GHD_refs:Doyon-Caux,Caux_2019}
and have been verified even experimentally \citep{Schemmer_2019}.
Moreover it has paved the way for interesting extensions and applications
\citep{GHD_refs:1,GHD_refs:2,GHD_refs:Bulchandani,GHD_refs:Piroli,GHD_refs:Collura,GHD_refs:Doyon-Caux,GHD_refs:Doyon-Caux-Dubail-Konik,GHD_refs:Doyon-Spohn1,GHD_refs:Doyon-Spohn2,GHD_refs:Doyon-Spohn3,Ruggiero2020,Bastianello_2020b,Vu2020},
as it essentially reduces the characterisation of macroscopic properties
resulting from the complicated dynamics of quantum many-body systems
to a relatively simple semiclassical description. At this point it
is worth to recognise the important role played by integrability in
understanding the physics of the quantum many-body problem in general,
especially in the field of out-of-equilibrium dynamics \citep{rev:Essler2016,rev:Calabrese2016,rev:Cazalilla2016,rev:Bernard2016,rev:Caux2016,rev:Vidmar2016,rev:Ilievski2016,rev:Langen2016,rev:Vasseur2016}.
In contrast to non-interacting models and perturbative techniques
that for many decades dominated any attempt to understand quantum
statistical mechanics, in recent years integrable models have emerged
as a powerful tool to unveil nontrivial and non-perturbative effects
of interactions that are nowhere seen in non-interacting models and
cannot always be studied satisfactorily by non-interacting approximations
or perturbative methods.

A rigorous proof or at least a better understanding of the mathematical
reasons that explain the emergence of GHD are highly desirable for
many reasons. First, even though the perfect agreement with numerical
simulations and experimental verification of its predictions leave
little space for doubts on its validity, GHD is still a conjecture
and a proof would enable us to identify possible exceptional special
cases. Second, the study of diffusion and even more of deviations
from standard diffusive behaviour, a subject of central interest in
quantum transport physics, can be based on next-to-leading order asymptotics
(beyond Euler scale) of the dynamics after inhomogeneous quenches
and therefore it would benefit essentially from exact results that
would be intermediate steps of a proof of GHD. Such results would
be useful for the verification of important results on diffusion that
have already been derived from extensions of GHD \citep{De_Nardis_2018,De_Nardis_2019a,De_Nardis_2019b,Ilievski_2018}.
Third, proving GHD can be an important step towards solving the more
general problem of quantum many-body dynamics in integrable systems
(and possibly beyond). In particular, it could pave the way for rigorously
proving the ``Quench Action'' method\citep{QA:Essler-Caux,rev:Caux2016},
which is the basis of GHD, or even extending it in such a way that
it can be used also for the study of intermediate dynamics. 

For these and many other reasons, several approaches to explain GHD
have been developed. Most of these have focused on the derivation
of a formula giving the expectation value of current operators in
generalised Gibbs ensembles, on which GHD is crucially based \citep{GHD_Doyon,Borsi2020,Pozsgay2020,Pozsgay_2020b,Spohn_2020}.
A general proof of this formula for quantum spin chains is now available
based on algebraic Bethe Ansatz techniques \citep{Borsi2020,Pozsgay_2020b}.
Another method could be based on form factor expansions \citep{Cort_s_Cubero_2019,CortesCubero2020,Cubero_2020,Essler_2020}
possibly by a suitable generalisation to the case of equilibrium states.

Starting with the case of effectively non-interacting dynamics presented
in this work, we aim to develop a different approach: based on an
exact analytical derivation of the asymptotics of local observables
after a partitioning quench, we wish to demonstrate the emergence
of the NESS dependent on the distance over time (ray) ratio and to
verify the GHD predictions for this protocol. For concreteness, we
consider here the Lieb-Liniger model \citep{Lieb_1963} in the Tonks-Girardeau
(``free fermion'') limit \citep{Tonks1936,Girardeau1960} and focus
on one of the simplest instances of this type of quench: the expansion
of a gas from one half to the entire confining box, a protocol that
is also known as ``geometric quench''. This protocol offers the
significant technical advantage that the expansion of the initial
state in the post-quench eigenstate basis is explicitly known for
any value of the interaction \citep{Mossel-Caux}. In the Tonks-Girardeau
limit the problem can be studied relatively easily using free fermion
techniques \citep{Collura2014}. Here we revisit this problem and
solve it using a different method that does not exploit its effectively
free character and reduction to the single particle problem. Our method
is based on complex analysis techniques that are rather robust, in
the sense that they are not too sensitive to model and protocol specific
details, for which reason it is promising for generalisation to the
far more interesting genuinely interacting case where GHD predictions
are non-trivial.

The paper is organised as follows. We first describe the mathematical
problem: the model and quench protocol, as well as our objectives
(sec. \ref{sec:Model}). In an interlude serving as a warm-up example,
we solve the single particle problem (sec. \ref{sec:1p}), stressing
the mathematical ideas that lead to the result and that will be later
generalised to the many particle problem. Passing to the main part,
we study the NESS for the Tonks-Girardeau gas expansion (sec. \ref{sec:TG}),
avoiding use of free fermion techniques. Lastly, we discuss the main
steps and essential ingredients of the method, evaluating the potential
for generalisation to the genuinely interacting case (sec. \ref{sec:Discussion}).

\section{Model and quench protocol\label{sec:Model}}

The Lieb-Liniger model describes a one-dimensional Bose gas with point-like
interactions. Its Hamiltonian is
\[
H_{LL}=\int_{-L/2}^{+L/2}\mathrm{d}x\,\left[-\Psi^{\dagger}(x)\partial_{x}^{2}\Psi(x)+c\,\Psi^{\dagger}(x)\Psi^{\dagger}(x)\Psi(x)\Psi(x)\right]
\]
where $c$ is the interaction strength and the particle mass has been
set to $m=1/2$. We assume periodic boundary conditions $\Psi(+L/2)=\Psi(-L/2)$.
 The particle number and momentum operators are respectively
\begin{align*}
N & =\int_{-L/2}^{+L/2}\mathrm{d}x\,\Psi^{\dagger}(x)\Psi(x)\\
P & =-\mathrm{i}\int_{-L/2}^{+L/2}\mathrm{d}x\,\Psi^{\dagger}(x)\partial_{x}\Psi(x)
\end{align*}

The Lieb-Liniger model is integrable i.e. its eigenstates are given
by the Bethe Ansatz for any value of the interaction $c$. In the
Tonks-Girardeau or hard-core boson limit $c\to\infty$ that we shall
consider in this work, the Bethe states and Bethe equations simplify
dramatically and the system becomes effectively equivalent to a system
of free fermions. This equivalence can be expressed also algebraically
through a Jordan-Wigner-like transformation relating the boson operators
$\Psi$ to hard-core boson operators $\Psi_{HB}$, which anti-commute
under exchange of their positions, except for coinciding positions.
In terms of these operators the Hamiltonian becomes non-interacting
\[
H_{TG}=\int_{-L/2}^{+L/2}\mathrm{d}x\,\left[-\Psi_{HB}^{\dagger}(x)\partial_{x}^{2}\Psi_{HB}(x)\right]
\]

The geometric quench problem we shall study is described as follows. We consider
a gas of $N$ particles, initially restricted in an interval of length
$L/2$, let us say in the right half of the system. We assume that
the system lies in an equilibrium state $\hat{\rho}_{0}$ of the (Lieb-Liniger
or Tonks-Girardeau) Hamiltonian restricted in this interval with periodic
boundary conditions $\Psi(+L/2)=\Psi(0)$. 

We do not impose any further conditions on the initial state: it
can be either the ground state $|\Phi_{\text{GS},0}\rangle$, an excited
state $|\Phi_{0}(\boldsymbol{\mu})\rangle$, a thermal state $\mathrm{e}^{-\beta H_{0}}$
or other diagonal ensemble of eigenstates, i.e. it is described by
a density matrix of the form
\[
\hat{\rho}_{0}=\mathcal{Z}_{0}^{-1}\sum_{\boldsymbol{\mu}:\text{ BA}_{0}}\rho_{0}(\boldsymbol{\mu})\frac{|\Phi_{0}(\boldsymbol{\mu})\rangle\langle\Phi_{0}(\boldsymbol{\mu})|}{\langle\Phi_{0}(\boldsymbol{\mu})|\Phi_{0}(\boldsymbol{\mu})\rangle}
\]
where $|\Phi_{0}(\boldsymbol{\lambda})\rangle$ are eigenstates of
$H_{0}$ with rapidities $\boldsymbol{\lambda}$ satisfying Bethe
equations corresponding to system size $L/2$ (BA$_{0}$)  and 
\[
\mathcal{Z}_{0}\equiv\sum_{\boldsymbol{\mu}:\text{ BA}_{0}}\rho_{0}(\boldsymbol{\mu})\langle\Phi_{0}(\boldsymbol{\mu})|\Phi_{0}(\boldsymbol{\mu})\rangle
\]
so that $\hat{\rho}_{0}$ satisfies the normalisation condition 
\begin{align*}
\text{Tr}\,\hat{\rho}_{0} & =\sum_{\boldsymbol{\mu}:\text{ BA}_{0}}\langle\Phi_{0}(\boldsymbol{\mu})|\hat{\rho}_{0}|\Phi_{0}(\boldsymbol{\mu})\rangle\\
 & =\mathcal{Z}_{0}^{-1}\sum_{\boldsymbol{\mu}:\text{ BA}_{0}}\rho_{0}(\boldsymbol{\mu})\langle\Phi_{0}(\boldsymbol{\mu})|\Phi_{0}(\boldsymbol{\mu})\rangle=1
\end{align*}
We then let the initial state evolve in time under the Hamiltonian
$H$ which is the same as $H_{0}$ except that it is defined in an
interval of length $L$, again with periodic boundary conditions $\Psi(+L/2)=\Psi(-L/2)$.

Our goal is to calculate the asymptotics of local observables $\hat{\mathcal{O}}(x,t)$
in the thermodynamic limit where both $N$ and $L$ tend to infinity
with fixed non-zero average density $N/L=n$ and at large times $t$
and distances $x$ keeping the ratio $x/t$ fixed. We do not impose
constraints on the observables $\hat{\mathcal{O}}$: they may be single-point
observables at position $x$ or multi-point correlation functions
at positions $x+r_{1},x+r_{2},\dots,x+r_{n}$ at fixed separations
$r_{i}$. The order of limits we consider is first the thermodynamic
limit and then the large time and distance limit. The quantity we
wish to calculate is therefore
\begin{equation}
\lim_{{x,t\to\infty\atop x/t=\xi:\text{ fixed}}}\lim_{{N,L\to\infty\atop n:\text{ fixed}}}\left\langle \hat{O}(x,t)\right\rangle =\lim_{{x,t\to\infty\atop x/t=\xi:\text{ fixed}}}\lim_{{N,L\to\infty\atop n:\text{ fixed}}}\text{Tr}\left\{ \hat{\rho}_{0}\,\mathrm{e}^{+\mathrm{i}Ht}\mathrm{e}^{-\mathrm{i}Px}\hat{O}\mathrm{e}^{+\mathrm{i}Px}\mathrm{e}^{-\mathrm{i}Ht}\right\} \label{eq:lim_obs}
\end{equation}
where $\hat{O}\equiv\hat{O}(0,0)$. 

We wish to show that a Non-Equilibrium Steady State (NESS) emerges
in this double limit and to provide a complete description of this
state. This means that our first objective is to show that in the
above limit the observable can be expressed as an expectation value
in a statistical ensemble $\hat{\rho}_{\text{NESS}}$ that is diagonal
in the post-quench eigenstate basis, i.e. of the form
\begin{equation}
\hat{\rho}_{\text{NESS}}=\mathcal{Z}_{\text{NESS}}^{-1}\sum_{\boldsymbol{\lambda}:\text{ BA}}\rho_{\text{\text{NESS}}}(\boldsymbol{\lambda})\frac{|\Phi(\boldsymbol{\lambda})\rangle\langle\Phi(\boldsymbol{\lambda})|}{\langle\Phi(\boldsymbol{\lambda})|\Phi(\boldsymbol{\lambda})\rangle}\label{eq:NESS-0}
\end{equation}
where $|\Phi(\boldsymbol{\lambda})\rangle$ are the eigenstates of
$H$ with rapidities $\boldsymbol{\lambda}$ satisfying the Bethe
equations corresponding to system size $L$ (BA). The state $\hat{\rho}_{\text{NESS}}$
will be different for each space-time ray, for which reason we will
denote it as $\hat{\rho}_{\text{\text{NESS},\ensuremath{\xi}}}$ to
show that it depends explicitly on the ratio $\xi=x/t$. Therefore
what we wish to show is that 
\begin{align}
\lim_{{x,t\to\infty\atop x/t=\xi:\text{ fixed}}}\lim_{{N,L\to\infty\atop n:\text{ fixed}}}\left\langle \hat{O}(x,t)\right\rangle  & =\lim_{{N,L\to\infty\atop n:\text{ fixed}}}\text{Tr}\left\{ \hat{\rho}_{\text{NESS},\xi}\,\hat{O}\right\} \label{eq:NESS}
\end{align}
where the expectation value is
\[
\text{Tr}\left\{ \hat{\rho}_{\text{\text{NESS},\ensuremath{\xi}}}\,\hat{O}\right\} =\mathcal{Z}_{\text{\text{NESS}}}^{-1}\sum_{\boldsymbol{\lambda}:\text{ BA}}\rho_{\text{\text{NESS}}}(\boldsymbol{\lambda};\xi)\langle\Phi(\boldsymbol{\lambda})|\hat{O}|\Phi(\boldsymbol{\lambda})\rangle
\]
The second objective is to determine this state in terms of the initial
state and the space-time ratio $\xi$, i.e. to express $\rho_{\text{\text{NESS}}}(\boldsymbol{\lambda};\xi)$
as a function of $\rho_{0}(\boldsymbol{\lambda})$ and $\xi$.

Expanding in the pre- and post-quench eigenstate bases, we can write
the time evolution of local observables in (\ref{eq:lim_obs}) as
\begin{align}
 & \left\langle \hat{O}(x,t)\right\rangle =\nonumber \\
 & =\mathcal{Z}_{0}^{-1}\text{Tr}\left\{ \left[\sum_{\boldsymbol{\mu}:\text{ BA}_{0}}\rho_{0}(\boldsymbol{\mu})\frac{|\Phi_{0}(\boldsymbol{\mu})\rangle\langle\Phi_{0}(\boldsymbol{\mu})|}{\langle\Phi_{0}(\boldsymbol{\mu})|\Phi_{0}(\boldsymbol{\mu})\rangle}\right]\left[\sum_{\boldsymbol{\lambda},\boldsymbol{\lambda}':\text{ BA}}\frac{|\Phi(\boldsymbol{\lambda}')\rangle\langle\Phi(\boldsymbol{\lambda}')|}{\langle\Phi(\boldsymbol{\lambda}')|\Phi(\boldsymbol{\lambda}')\rangle}\mathrm{e}^{+\mathrm{i}Ht}\mathrm{e}^{-\mathrm{i}Px}\hat{O}\mathrm{e}^{+\mathrm{i}Px}\mathrm{e}^{-\mathrm{i}Ht}\frac{|\Phi(\boldsymbol{\lambda})\rangle\langle\Phi(\boldsymbol{\lambda})|}{\langle\Phi(\boldsymbol{\lambda})|\Phi(\boldsymbol{\lambda})\rangle}\right]\right\} \nonumber \\
 & =\mathcal{Z}_{0}^{-1}\sum_{\boldsymbol{\mu}:\text{ BA}_{0}}\sum_{\boldsymbol{\lambda},\boldsymbol{\lambda}':\text{ BA}}\rho_{0}(\boldsymbol{\mu})\mathrm{e}^{+\mathrm{i}[P(\boldsymbol{\lambda})-P(\boldsymbol{\lambda}')]x-\mathrm{i}[E(\boldsymbol{\lambda})-E(\boldsymbol{\lambda}')]t}\frac{\mathcal{M}^{*}(\boldsymbol{\lambda}';\boldsymbol{\mu})\mathcal{M}(\boldsymbol{\lambda};\boldsymbol{\mu})}{\mathcal{N}_{0}(\boldsymbol{\mu})\mathcal{N}(\boldsymbol{\lambda}')\mathcal{N}(\boldsymbol{\lambda})}\langle\Phi(\boldsymbol{\lambda}')|\hat{O}|\Phi(\boldsymbol{\lambda})\rangle\label{eq:time-evol-gen}
\end{align}
where we defined the norms of post- and pre-quench eigenstates as
\begin{align*}
\mathcal{N}(\boldsymbol{\lambda}) & \equiv\langle\Phi(\boldsymbol{\lambda})|\Phi(\boldsymbol{\lambda})\rangle\\
\mathcal{N}_{0}(\boldsymbol{\lambda}) & \equiv\langle\Phi_{0}(\boldsymbol{\lambda})|\Phi_{0}(\boldsymbol{\lambda})\rangle
\end{align*}
since in their standard form the Bethe Ansatz eigenstates are not automatically 
normalised, and their overlaps as
\begin{align*}
\mathcal{M}(\boldsymbol{\lambda};\boldsymbol{\mu}) & \equiv\langle\Phi(\boldsymbol{\lambda})|\Phi_{0}(\boldsymbol{\mu})\rangle
\end{align*}

Since we expect that the result we wish to show is valid for a general
local observable and a general initial state diagonal in the pre-quench
basis, the proof should not generally rely on special properties of
the matrix elements $\langle\Phi(\boldsymbol{\lambda}')|\hat{O}|\Phi(\boldsymbol{\lambda})\rangle$
or of the amplitudes $\rho_{0}(\boldsymbol{\mu})$. Motivated by this
observation, the strategy we shall pursue is to analyse the asymptotics
of any pre-quench eigenstate time evolved under the post-quench Hamiltonian.
More specifically, we shall consider the quantity 
\begin{align*}
\mathrm{e}^{+\mathrm{i}Px-\mathrm{i}Ht}|\Phi_{0}(\boldsymbol{\mu})\rangle & =\sum_{\boldsymbol{\lambda}:\text{ BA}}\mathrm{e}^{+\mathrm{i}P(\boldsymbol{\lambda})x-\mathrm{i}E(\boldsymbol{\lambda})t}\frac{|\Phi(\boldsymbol{\lambda})\rangle\langle\Phi(\boldsymbol{\lambda})|}{\langle\Phi(\boldsymbol{\lambda})|\Phi(\boldsymbol{\lambda})\rangle}|\Phi_{0}(\boldsymbol{\mu})\rangle\\
 & =\sum_{\boldsymbol{\lambda}:\text{ BA}}\mathrm{e}^{+\mathrm{i}P(\boldsymbol{\lambda})x-\mathrm{i}E(\boldsymbol{\lambda})t}\frac{\mathcal{M}(\boldsymbol{\lambda};\boldsymbol{\mu})}{\mathcal{N}(\boldsymbol{\lambda})}|\Phi(\boldsymbol{\lambda})\rangle
\end{align*}
where we have introduced a resolution of the identity in the post-quench
basis. What we aim to show is that, as long as coordinate space wave-functions
are concerned, the following asymptotic relation holds
\begin{align}
\lim_{{x,t\to\infty\atop x/t=\xi:\text{ fixed}}}\lim_{{N,L\to\infty\atop n:\text{ fixed}}}\mathrm{e}^{+\mathrm{i}Px-\mathrm{i}Ht}|\Phi_{0}(\boldsymbol{\mu})\rangle & \sim\lim_{N\to\infty}\mathcal{F}(\boldsymbol{\mu};x/t)\mathrm{e}^{+\mathrm{i}P(\boldsymbol{\mu})x-\mathrm{i}E(\boldsymbol{\mu})t}|\Phi(\boldsymbol{\mu})\rangle\label{eq:CONJ}
\end{align}
that is, any pre-quench eigenstate time-evolved by $t$ and space-translated
by $x$ can be replaced in the thermodynamics limit by a post-quench
eigenstate with the same rapidity distribution. The equivalence in
the above equation is meant to hold asymptotically in the double limit,
as long as the operator or state that acts from the left is local. 

If this equation holds then (\ref{eq:time-evol-gen}) gives
\[
\lim_{{x,t\to\infty\atop x/t=\xi:\text{ fixed}}}\lim_{{N,L\to\infty\atop n:\text{ fixed}}}\left\langle \hat{O}(x,t)\right\rangle =\lim_{{N,L\to\infty\atop n:\text{ fixed}}}\mathcal{Z}_{0}^{-1}\sum_{\boldsymbol{\mu}:\text{ BA}}\frac{\rho_{0}(\boldsymbol{\mu})}{\mathcal{N}_{0}(\boldsymbol{\mu})}\left|\mathcal{F}(\boldsymbol{\mu};x/t)\right|^{2}\langle\Phi(\boldsymbol{\mu})|\hat{O}|\Phi(\boldsymbol{\mu})\rangle
\]
which is indeed of the form (\ref{eq:NESS}) with the NESS given by
\begin{align}
\frac{\rho_{\text{NESS}}(\boldsymbol{\lambda};x/t)}{\mathcal{Z}_{\text{\text{NESS}}}} & =\frac{\rho_{0}(\boldsymbol{\lambda})}{\mathcal{Z}_{0}\mathcal{N}_{0}(\boldsymbol{\lambda})}\left|\mathcal{F}(\boldsymbol{\lambda};x/t)\right|^{2}\label{eq:NESS-GHD}
\end{align}
therefore our goal will have been achieved.

\section{Single particle problem\label{sec:1p}}

Before we study the problem of derivation of the NESS in the Tonks-Girardeau
limit of the Lieb-Liniger model, it is instructive to study the single
particle problem first, which offers a pedagogic exposition of the
necessary mathematical tools. The method presented here is essentially
the same as that of \citep{defect} applied to a similar problem.

We consider a particle in a box of length $L/2$, initially lying
at one of the eigenstates $|\Phi_{0}(q)\rangle$ of the Hamiltonian
$H_{0}=-\partial_{x}^{2}$ with periodic boundary conditions at the
points $x=0$ and $L/2$. At $t=0$ we extend the length of the box
from $L/2$ to $L$ and let the initial state evolve under the Hamiltonian
$H=-\partial_{x}^{2}$ with periodic boundary conditions at the points
$x=-L/2$ and $+L/2$. Our strategy for the derivation of the NESS
is based on the study of the asymptotics of the time evolved coordinate
space wave-function of an arbitrary pre-quench eigenstate $|\Phi_{0}(q)\rangle$. 

The coordinate space wave-functions of pre-quench eigenstates are
\begin{align*}
\langle x|\Phi_{0}(q)\rangle & =\mathrm{e}^{\mathrm{i}qx}\Theta(x)
\end{align*}
with the quantised momenta $q$ given by
\begin{equation}
q=2\pi n_{0}/(L/2)=4\pi n_{0}/L,\quad n_{0}\in\mathbb{Z}\label{eq:q-cond0}
\end{equation}
while the post-quench eigenstates are
\begin{align*}
\langle x|\Phi(k)\rangle & =\mathrm{e}^{\mathrm{i}kx}
\end{align*}
with the quantised momenta $k$ given by 
\begin{equation}
k=2\pi n/L,\quad n\in\mathbb{Z}\label{eq:q-cond}
\end{equation}
The norms of pre-quench eigenstates $\mathcal{N}_{0}(q)$ and of post-quench
eigenstates $\mathcal{N}(k)$ are 
\begin{align*}
\mathcal{N}_{0}(q) & =L/2\\
\mathcal{N}(k) & =L
\end{align*}
respectively.

We are interested in the asymptotic form of the wavefunction in the
infinite length limit $L\to\infty$ at large distance $x$ and time
$t$ with fixed ratio $\xi=x/t$. More specifically, we will calculate
the asymptotics of the quantity
\begin{equation}
\mathcal{K}(q;z;x,t)\equiv\langle z|\mathrm{e}^{+\mathrm{i}Px-\mathrm{i}Ht}|\Phi_{0}(q)\rangle\label{eq:K}
\end{equation}
where $P=-{\rm i}\partial_{x}$ is the momentum operator, which generates
space translations. Knowing $\mathcal{K}(q;z;x,t)$ we can derive
the asymptotics of any local observable in any initial state that
is diagonal in the pre-quench basis. 

Introducing a resolution of the identity in the post-quench basis
we have
\begin{align}
\mathrm{e}^{+\mathrm{i}Px-\mathrm{i}Ht}|\Phi_{0}(q)\rangle & =\sum_{{k=2\pi n/L\atop n\in\mathbb{Z}}}\mathrm{e}^{+\mathrm{i}P(k)x-\mathrm{i}E(k)t}\frac{|\Phi(k)\rangle\langle\Phi(k)|}{\langle\Phi(k)|\Phi(k)\rangle}|\Phi_{0}(q)\rangle\nonumber \\
 & =\sum_{{k=2\pi n/L\atop n\in\mathbb{Z}}}\mathrm{e}^{+\mathrm{i}P(k)x-\mathrm{i}E(k)t}\frac{\mathcal{M}(k;q)}{\mathcal{N}(k)}|\Phi(k)\rangle\label{eq:K1}
\end{align}
where 
\begin{align*}
\mathcal{M}(k;q) & \equiv\langle\Phi(k)|\Phi_{0}(q)\rangle\\
\mathcal{N}(k) & \equiv\langle\Phi(k)|\Phi(k)\rangle
\end{align*}
are the overlaps between pre- and post-quench eigenstates and the
norm square of the latter respectively, and 
\begin{align}
E(k) & =k^{2}\label{eq:TGe}\\
P(k) & =k\label{eq:TGp}
\end{align}
are the energy and momentum eigenvalues.

From the above we can easily calculate the overlaps between pre- and
post-quench eigenstates
\begin{align*}
\mathcal{M}(k;q) & =\int_{0}^{L/2}\mathrm{d}x\;\mathrm{e}^{\mathrm{i}(q-k)x}\\
 & =\frac{\mathrm{i}}{k-q}\left({\rm e}^{{\rm i}(q-k)L/2}-1\right)
\end{align*}
Using the pre- and post-quench quantisation conditions (\ref{eq:q-cond0})
and (\ref{eq:q-cond}), we find that the overlaps acquire a different
form as functions of the momenta in the even and odd $n$ sectors
\begin{align}
\mathcal{M}(k;q) & =\begin{cases}
(L/2)\delta_{n,2n_{0}}=(L/2)\delta_{k,q} & \text{ if }n=\frac{kL}{2\pi}\text{ is even}\\
-2\mathrm{i}(L/2\pi)/(n-2n_{0})=-2\mathrm{i}/(k-q) & \text{ if }n\text{ is odd}
\end{cases}\label{eq:TGoverlaps}
\end{align}
This means that in the thermodynamic limit $L\to\infty$ we should
integrate separately in the two sectors. It is convenient to redefine
the overlap $\mathcal{M}(k;q)$ as a function of an additional index
$s$ taking values $+1$ if $n$ is even and $-1$ if it is odd. Note
that the index $s$ is itself a function of the discrete momentum
variable $k$ which is a solution of the quantisation conditions.
But we could also see it as an independent discrete variable and consider
$k$ as a continuous variable, since $s$ is the only information
about the discreteness of the momentum that might be relevant in the
thermodynamic limit. We therefore write 
\begin{align*}
\mathcal{M}(k;q) & \eqqcolon\begin{cases}
\mathcal{M}_{+}(k;q) & \text{ if }n=\frac{kL}{2\pi}\text{ is even}\\
\mathcal{M}_{-}(k;q) & \text{ if }n\text{ is odd}
\end{cases}
\end{align*}
and from (\ref{eq:TGoverlaps})
\begin{align*}
\mathcal{M}_{s}(k;q) & =\begin{cases}
(L/2)\delta_{k,q} & \text{ if }s=+1\\
-2\mathrm{i}/(k-q) & \text{ if }s=-1
\end{cases}
\end{align*}

Moreover in the odd sector $\mathcal{M}_{-}(k;q)$ develops a pole
at $k=q$, a value that is of course absent in the discrete set of
$k$ values for odd $n$. This implies that the corresponding sum
cannot be directly expressed as an integral along the real $k$ axis,
but the correct prescription for how to avoid the pole must be specified.
This can be done using a standard complex analysis trick, as explained
below.

Substituting (\ref{eq:TGoverlaps}) into (\ref{eq:K1}) and evaluating
the even sector sum we obtain
\begin{align}
\mathcal{K}(q;z;x,t) & =\sum_{s=\pm}\sum_{{k=2\pi n_{s}/L\atop n_{\pm}:\text{ even/odd}}}\mathrm{e}^{+\mathrm{i}P(k)x-\mathrm{i}E(k)t}\frac{\mathcal{M}_{s}(k;q)}{\mathcal{N}(k)}\langle z|\Phi(k)\rangle\label{eq:K0}\\
 & =\frac{1}{2}\mathrm{e}^{+\mathrm{i}P(q)x-\mathrm{i}E(q)t}\langle z|\Phi(q)\rangle+\frac{1}{L}\sum_{{k=2\pi n/L\atop n\text{ odd}}}\mathrm{e}^{+\mathrm{i}P(k)x-\mathrm{i}E(k)t}\left(\frac{-2\mathrm{i}}{k-q}\right)\langle z|\Phi(k)\rangle\label{eq:K2}
\end{align}
In order to write the odd sector sum $S_{-}$ as an integral in the
infinite system size limit $L\to\infty$, we first introduce a function
of $k$ that has simple poles of residue $1/(2\pi\mathrm{i})$ at
the points $k=2\pi n/L$ for all odd integers $n=2\ell+1$ so that
we can write each term of the sum as an integral over $k$ along a
small circle $\mathcal{C}_{\ell}=C(2\pi(2\ell+1)/L,\epsilon)$ around
each of these points. The suitable function is 
\begin{equation}
F_{-}(k)\equiv-\frac{L}{4\pi}\frac{1}{\mathrm{e}^{\mathrm{i}kL/2}+1}\label{eq:F-}
\end{equation}
and we obtain
\begin{align*}
S_{-} & \equiv\frac{1}{L}\sum_{{k=2\pi n/L\atop n\text{ odd}}}\mathrm{e}^{+\mathrm{i}P(k)x-\mathrm{i}E(k)t}\left(\frac{-2\mathrm{i}}{k-q}\right)\langle z|\Phi(k)\rangle\\
 & =\frac{1}{L}\sum_{{k=2\pi(2\ell+1)/L\atop \ell\in\mathbb{Z}}}\mathrm{e}^{+\mathrm{i}P(k)x-\mathrm{i}E(k)t}\left(\frac{-2\mathrm{i}}{k-q}\right)\langle z|\Phi(k)\rangle\\
 & =-\sum_{\ell\in\mathbb{Z}}\ointop\limits _{\mathcal{C}_{\ell}}\frac{\mathrm{d}k}{4\pi}\,\frac{1}{\mathrm{e}^{\mathrm{i}kL/2}+1}\mathrm{e}^{+\mathrm{i}P(k)x-\mathrm{i}E(k)t}\left(\frac{-2\mathrm{i}}{k-q}\right)\langle z|\Phi(k)\rangle
\end{align*}
Next, using the analyticity of the integrand everywhere else on the
real $k$ axis except at $k=q$ where it is singular, we merge the
circles into another contour composed of two straight lines $\mathcal{C}_{\pm}=\mathbb{R}\pm\mathrm{i}\epsilon$
running just above and just below the real $k$ axis and a small circle
$\mathcal{C}_{p}=C(q,\epsilon)$ in order to subtract the contribution
of the pole at $k=q$
\begin{align*}
S_{-} & =-\int\limits _{\mathcal{C}_{-}-\mathcal{C}_{+}-\mathcal{C}_{p}}\frac{\mathrm{d}k}{4\pi}\,\frac{1}{\mathrm{e}^{\mathrm{i}kL/2}+1}\mathrm{e}^{+\mathrm{i}P(k)x-\mathrm{i}E(k)t}\left(\frac{-2\mathrm{i}}{k-q}\right)\langle z|\Phi(k)\rangle\\
 & =-\left(\intop_{-\infty-\mathrm{i}\epsilon}^{+\infty-\mathrm{i}\epsilon}-\intop_{-\infty+\mathrm{i}\epsilon}^{+\infty+\mathrm{i}\epsilon}\;\right)\frac{\mathrm{d}k}{4\pi}\,\frac{1}{\mathrm{e}^{\mathrm{i}kL/2}+1}\mathrm{e}^{+\mathrm{i}P(k)x-\mathrm{i}E(k)t}\left(\frac{-2\mathrm{i}}{k-q}\right)\langle z|\Phi(k)\rangle+\\
 & +2\pi\mathrm{i}\frac{1}{4\pi}\,\frac{1}{\mathrm{e}^{\mathrm{i}qL/2}+1}\mathrm{e}^{+\mathrm{i}P(q)x-\mathrm{i}E(q)t}\langle z|\Phi(q)\rangle\;\underset{k=q}{\text{Res}}\left(\frac{-2\mathrm{i}}{k-q}\right)
\end{align*}
We now evaluate the contribution of the pole at $k=q$, which is a
simple pole due to the factor $1/(k-q)$ since the rest of the integrand
is analytic at that point. The factor $1/\left(\mathrm{e}^{\mathrm{i}qL/2}+1\right)$
equals $1/2$ for any value of $q$ because, by satisfying the pre-quench
quantisation conditions, $q$ is equal to $4\pi n_{0}/L$ with $n_{0}\in\mathbb{Z}$.
Substituting into the last equation we obtain
\begin{align*}
S_{-} & =-\left(\intop_{-\infty-\mathrm{i}\epsilon}^{+\infty-\mathrm{i}\epsilon}-\intop_{-\infty+\mathrm{i}\epsilon}^{+\infty+\mathrm{i}\epsilon}\;\right)\frac{\mathrm{d}k}{4\pi}\,\frac{1}{\mathrm{e}^{\mathrm{i}kL/2}+1}\mathrm{e}^{+\mathrm{i}P(k)x-\mathrm{i}E(k)t}\left(\frac{-2\mathrm{i}}{k-q}\right)\langle z|\Phi(k)\rangle+\\
 & +\frac{1}{2}\mathrm{e}^{+\mathrm{i}P(q)x-\mathrm{i}E(q)t}\langle z|\Phi(q)\rangle
\end{align*}

Lastly, we take the infinite system size limit $L\to\infty$. We notice
that 
\[
\lim_{L\to\infty}\frac{1}{\mathrm{e}^{\mathrm{i}kL/2}+1}=\begin{cases}
1 & \text{ if Im}(k)>0\\
0 & \text{ if Im}(k)<0
\end{cases}
\]
Therefore the integral along the line below the real axis vanishes
in this limit, while the remaining integral along the line above the
real axis gives
\begin{align*}
\lim_{L\to\infty}S_{-} & =\intop_{-\infty+\mathrm{i}\epsilon}^{+\infty+\mathrm{i}\epsilon}\frac{\mathrm{d}k}{2\pi}\,\mathrm{e}^{+\mathrm{i}P(k)x-\mathrm{i}E(k)t}\left(\frac{-\mathrm{i}}{k-q}\right)\langle z|\Phi(k)\rangle+\\
 & +\frac{1}{2}\mathrm{e}^{+\mathrm{i}P(q)x-\mathrm{i}E(q)t}\langle z|\Phi(q)\rangle
\end{align*}
Substituting the final result into (\ref{eq:K0}), we obtain
\begin{equation}
\lim_{L\to\infty}\mathcal{K}(q;z;x,t)=\mathrm{e}^{+\mathrm{i}P(q)x-\mathrm{i}E(q)t}\langle z|\Phi(q)\rangle+\intop_{-\infty+\mathrm{i}\epsilon}^{+\infty+\mathrm{i}\epsilon}\frac{\mathrm{d}k}{2\pi}\,\mathrm{e}^{+\mathrm{i}P(k)x-\mathrm{i}E(k)t}\left(\frac{-\mathrm{i}}{k-q}\right)\langle z|\Phi(k)\rangle\label{eq:TGK}
\end{equation}

Overall what we have achieved is to calculate the $L\to\infty$ limit
of the sum over all modes in $\mathcal{K}(q;z;x,t)$ as an expression
involving an integral on a line parallel to the real axis with the
precise prescription on how to avoid the momentum pole at $k=q$.

Having taken the thermodynamic limit, we now proceed to derive the
asymptotics at large $x,t$ keeping the ratio $x/t\equiv\xi$ fixed.
We focus on the second term that involves the integral. The asymptotics
of such integrals can be found easily using another complex analysis
trick. Provided that the real part of the exponent of $\mathrm{e}^{+\mathrm{i}P(k)x-\mathrm{i}E(k)t}=\mathrm{e}^{+\mathrm{i}\left(P(k)\xi-\mathrm{i}E(k)\right)t}$
is negative for all $k$ along the integration contour, the integral
decays when $x,t\to\infty$. If it is non-negative for part of the
contour, then in order to calculate its asymptotics it is sufficient
to deform the contour into the region of the complex plane where it
becomes negative. In doing so we will have to take into account if
the contour crosses any singularity of the integrand, in which case
we will have to subtract its contribution. The wave-function $\langle z|\Phi(k)\rangle$,
being a simple plane wave, is analytic in $k$, so the only singularity
that may be crossed is the pole at $k=q$. 

The condition that the exponent has a negative real part gives
\begin{align*}
\text{Re}\left[\mathrm{i}P(k+\mathrm{i}\epsilon)x-\mathrm{i}E(k+\mathrm{i}\epsilon)t\right] & =\\
\text{Re}\left[\mathrm{i}\left(P(k)+\mathrm{i}\epsilon P'(k)\right)x-\mathrm{i}\left(E(k)+\mathrm{i}\epsilon E'(k)\right)t\right] & =\\
\left(-P'(k)\xi+E'(k)\right)\epsilon t & <0
\end{align*}
where we made use of the analyticity of $P(k)$ and $E(k)$ on the
real axis. We therefore find that for all $k$ such that 
\[
\frac{E'(k)}{P'(k)}<\xi
\]
(which in the present case means for $2k<\xi$) we keep the original
positive shift $\epsilon>0$ and the integral in this interval vanishes
as $x,t\to\infty$. On the other hand, in the interval where $k$
satisfies the inequality 
\[
\frac{E'(k)}{P'(k)}>\xi
\]
(i.e. for $2k>\xi$ in the present case) we should deform the contour
from above the real axis to below so that $\epsilon<0$ and the condition
is met. If the value $k=q$ lies within the interval where the latter
inequality holds, then the momentum pole at $k=q$ is crossed during
the deformation and its residue should be subtracted. As the deformed
integral vanishes, the asymptotics is given precisely by the pole
contribution. Note that in the above inequalities the quantity
\[
v(k)\equiv\frac{E'(k)}{P'(k)}=\frac{\mathrm{d}E}{\mathrm{d}P}
\]
can be recognised as the group velocity of the free particle excitations
as given by their dispersion relation.

In the simple case where the group velocity $v(k)$ is a monotonically
increasing function of the momentum $k$ (as in the present case where
$v(k)=2k$), the inequalities $v(k)\lessgtr\xi$ reduce to $k\lessgtr k_{*}(\xi)$
where the threshold value $k_{*}(\xi)$ is given by the equation 
\[
v(k_{*}(\xi))=\xi
\]
whose solution in the present case is $k_{*}(\xi)=\xi/2$.

Following the above observations, we write
\begin{align}
 & \intop_{-\infty+\mathrm{i}\epsilon}^{+\infty+\mathrm{i}\epsilon}\frac{\mathrm{d}k}{2\pi}\,\mathrm{e}^{+\mathrm{i}P(k)x-\mathrm{i}E(k)t}\left(\frac{-\mathrm{i}}{k-q}\right)\langle z|\Phi(k)\rangle\nonumber \\
 & =\intop_{-\infty+\mathrm{i}\epsilon}^{+\infty+\mathrm{i}\epsilon}\frac{\mathrm{d}k}{2\pi}\,\Theta(\xi-v(k))\mathrm{e}^{+\mathrm{i}P(k)x-\mathrm{i}E(k)t}\left(\frac{-\mathrm{i}}{k-q}\right)\langle z|\Phi(k)\rangle\nonumber \\
 & +\intop_{-\infty-\mathrm{i}\epsilon}^{+\infty-\mathrm{i}\epsilon}\frac{\mathrm{d}k}{2\pi}\,\Theta(v(k)-\xi)\mathrm{e}^{+\mathrm{i}P(k)x-\mathrm{i}E(k)t}\left(\frac{-\mathrm{i}}{k-q}\right)\langle z|\Phi(k)\rangle\nonumber \\
 & +\intop_{k_{*}(\xi)+\mathrm{i}\epsilon}^{k_{*}(\xi)-\mathrm{i}\epsilon}\frac{\mathrm{d}k}{2\pi}\,\mathrm{e}^{+\mathrm{i}P(k)x-\mathrm{i}E(k)t}\left(\frac{-\mathrm{i}}{k-q}\right)\langle z|\Phi(k)\rangle\nonumber \\
 & -2\pi\mathrm{i}\frac{1}{2\pi}\Theta(v(q)-\xi)\mathrm{e}^{+\mathrm{i}P(q)x-\mathrm{i}E(q)t}\langle z|\Phi(q)\rangle\underset{k=q}{\text{Res}}\left(\frac{-\mathrm{i}}{k-q}\right)\label{eq:trick2}
\end{align}
Note that the integrals of the first and second line of (\ref{eq:trick2})
decay exponentially with $x,t$, while that of the third line decays
algebraically and is the one that determines the scaling of decaying
corrections to the asymptotics. Therefore we obtain
\begin{align*}
 & \intop_{-\infty+\mathrm{i}\epsilon}^{+\infty+\mathrm{i}\epsilon}\frac{\mathrm{d}k}{2\pi}\,\mathrm{e}^{+\mathrm{i}P(k)x-\mathrm{i}E(k)t}\left(\frac{-\mathrm{i}}{k-q}\right)\langle z|\Phi(k)\rangle\\
 & \stackrel[x/t=\xi\text{ fixed}]{|x|,t\to\infty}{\longrightarrow}-\Theta(v(q)-\xi)\mathrm{e}^{+\mathrm{i}P(q)x-\mathrm{i}E(q)t}\langle z|\Phi(q)\rangle
\end{align*}

Substituting our result into (\ref{eq:TGK}) we find the rather simple
and elegant formula
\[
\lim_{L\to\infty}\mathcal{K}(q;z;x,t)\stackrel[x/t=\xi\text{ fixed}]{|x|,t\to\infty}{\longrightarrow}\Theta(\xi-v(q))\mathrm{e}^{+\mathrm{i}P(q)x-\mathrm{i}E(q)t}\langle z|\Phi(q)\rangle
\]
which can also be written as
\[
\lim_{{x,t\to\infty\atop x/t=\xi:\text{ fixed}}}\lim_{L\to\infty}\mathrm{e}^{-\mathrm{i}P(q)x+\mathrm{i}E(q)t}\mathcal{K}(q;z;x,t)=\Theta(\xi-v(q))\langle z|\Phi(q)\rangle
\]
which concludes our calculation.

This formula directly leads to the conclusion that for a general initial
state that is diagonal in the pre-quench basis and for a general local
observable $\hat{\mathcal{O}}$, the large distance and time asymptotics
in the thermodynamic limit is given by a NESS that is diagonal in
the post-quench basis. Indeed we have
\begin{align*}
\lim_{{x,t\to\infty\atop x/t=\xi:\text{ fixed}}}\lim_{L\to\infty}\left\langle \hat{O}(x,t)\right\rangle  & =\lim_{{x,t\to\infty\atop x/t=\xi:\text{ fixed}}}\lim_{L\to\infty}\text{Tr}\left\{ \hat{\rho}_{0}\mathrm{e}^{+\mathrm{i}Ht}\mathrm{e}^{-\mathrm{i}Px}\hat{O}\mathrm{e}^{+\mathrm{i}Px}\mathrm{e}^{-\mathrm{i}Ht}\right\} \\
 & =\lim_{{x,t\to\infty\atop x/t=\xi:\text{ fixed}}}\lim_{L\to\infty}\mathcal{Z}_{0}^{-1}\sum_{{q=2\pi n_{0}/(L/2)\atop n_{0}\in\mathbb{Z}}}\int_{-L/2}^{+L/2}\mathrm{d}y\mathrm{d}y'\,\mathcal{N}_{0}^{-1}(q)\rho_{0}(q)\mathcal{K}^{*}(q;y';x,t)\mathcal{K}(q;y;x,t)\langle y'|\hat{O}|y\rangle\\
 & =\mathcal{Z}_{0}^{-1}\int_{-\infty}^{+\infty}\frac{\mathrm{d}q}{2\pi}\rho_{0}(q)\Theta(\xi-v(q))\int_{-\infty}^{+\infty}\mathrm{d}y\mathrm{d}y'\,\langle\Phi(q)|y'\rangle\langle y'|\hat{O}|y\rangle\langle y|\Phi(q)\rangle\\
 & =\mathcal{Z}_{0}^{-1}\int_{-\infty}^{+\infty}\frac{\mathrm{d}q}{2\pi}\rho_{0}(q)\Theta(\xi-v(q))\langle\Phi(q)|\hat{O}|\Phi(q)\rangle
\end{align*}
from which we see that the NESS is described by the momentum distribution
\[
\rho_{\text{NESS}}(q;\xi)\propto\rho_{0}(q)\Theta(\xi-v(q))
\]

Note that $\rho_{\text{NESS}}(q;\xi)$ is space-time dependent through
the ray parameter $\xi=x/t$, being equal to the initial momentum
distribution $\rho_{0}(q)$ for all momenta $q$ such that the corresponding
group velocity is $v(q)<\xi$ and to zero otherwise. This is what
is expected on the basis of the simple semiclassical picture of ballistically
moving quasiparticles: at any ray $\xi$ the corresponding NESS is
the result of the mixing of quasiparticles originating from either
the left or the right half side with the corresponding momentum distribution
i.e. $\rho_{\text{NESS}}(q;\xi)\propto\rho_{0L}(q)\Theta(v(q)-\xi)+\rho_{0R}(q)\Theta(\xi-v(q))$.
In the present case $\rho_{0L}(q)=0$ and $\rho_{0R}(q)=\rho_{0}(q)$,
since initially all particles are on the right side and the left side
is empty. The latter formula which extends our result for the gas
expansion to the more general case of gas mixing, can be derived straightforwardly
following the same steps as above for a suitable initial state. 

\section{Tonks-Girardeau limit of the Lieb-Liniger model\label{sec:TG}}

Having clearly explained the mathematical tools in the single case
particle case, we can now proceed to the many particle problem. As
already explained, in the Tonks-Girardeau limit the model is effectively
equivalent to non-interacting fermions. Based on the exact mapping
to free fermions, in principle the calculation of any observable reduces
through Wick's theorem to the calculation of the fermionic two-point
correlation function $\langle\Psi_{F}^{\dagger}(x,t)\Psi_{F}(x+r,t)\rangle$,
i.e. essentially to the single-particle problem we have just solved,
apart from complications arising from the non-trivial expression of
bosonic observables in terms of hard-core boson or fermion operators.
For the purpose of generalising our approach to the genuinely interacting
case, however, we will perform the calculation without taking advantage
of Wick's theorem, confronting the mathematical challenges of the
many-body problem.

We start by writing the exact coordinate space eigenfunctions, which
are
\begin{equation}
\langle\boldsymbol{z}|\Phi(\boldsymbol{\lambda})\rangle=\frac{1}{\sqrt{N!}}\det_{i,j}\left[\exp\left(\mathrm{i}\lambda_{j}x_{i}\right)\right]\prod_{j>i}\text{sign}(x_{j}-x_{i})\label{eq:TG_wf}
\end{equation}
where the rapidities $\boldsymbol{\lambda}$ in the system of length
$L$ are quantised following the equations 
\begin{equation}
\exp\left(\mathrm{i}\lambda_{j}L\right)=(-1)^{N-1}\qquad\text{for all }j=1,2,\dots,N\label{eq:TG_BA}
\end{equation}
These are nothing but the Bethe equations, which in this limit reduce
to the above decoupled form, and can be solved easily to give
\begin{equation}
\lambda_{j}=\begin{cases}
\frac{2\pi}{L}n_{j} & \text{ for }N\text{ odd}\\
\frac{2\pi}{L}\left(n_{j}+\frac{1}{2}\right) & \text{ for }N\text{ even}
\end{cases},\quad n_{j}\in\mathbb{Z}\label{eq:TG_BA_sol}
\end{equation}
Note that both the eigenfunctions and the rapidity eigenvalues take
the same form as for a system of free fermions, apart from the odd-even
selection rule on the values of the rapidities and the presence of
the sign product in (\ref{eq:TG_wf}), which guarantees symmetry of
the wavefunctions under exchange of the bosonic particles and accounts
for the difference between hard-core bosons and fermions. 

The energy and momentum eigenvalues corresponding to an eigenstate
with rapidities $\boldsymbol{\lambda}$ are respectively 
\begin{align*}
E(\boldsymbol{\lambda}) & =\sum_{i=1}^{N}e(\lambda_{i})=\sum_{i=1}^{N}\lambda_{i}^{2}\\
P(\boldsymbol{\lambda}) & =\sum_{i=1}^{N}p(\lambda_{i})=\sum_{i=1}^{N}\lambda_{i}
\end{align*}

The quantisation conditions for the half system are 
\begin{equation}
\exp\left(\mathrm{i}\mu_{j}L/2\right)=(-1)^{N-1}\qquad\text{for all }j=1,2,\dots,N\label{eq:TG_BA0}
\end{equation}
with solutions
\[
\mu_{j}=\begin{cases}
\frac{4\pi}{L}n_{j} & \text{ for }N\text{ odd}\\
\frac{4\pi}{L}\left(n_{j}+\frac{1}{2}\right) & \text{ for }N\text{ even}
\end{cases},\quad n_{j}\in\mathbb{Z}
\]

As before, we will focus on the derivation of the asymptotics of pre-quench
eigenstates $|\Phi_{0}(\boldsymbol{\mu})\rangle$ time evolved under
the Tonks-Girardeau Hamiltonian $H_{TG}$ and projected on a local
basis. We therefore wish to calculate the asymptotics of the quantity
\begin{equation}
\mathcal{K}(\boldsymbol{\mu};\boldsymbol{z};x,t)\equiv\langle\boldsymbol{z}|\mathrm{e}^{+\mathrm{i}Px-\mathrm{i}Ht}|\Phi_{0}(\boldsymbol{\mu})\rangle\label{eq:Kdef}
\end{equation}
Introducing a resolution of the identity in the post-quench basis
we have 
\begin{align}
\mathrm{e}^{+\mathrm{i}Px-\mathrm{i}Ht}|\Phi_{0}(\boldsymbol{\mu})\rangle & =\sum_{\boldsymbol{\lambda}:\text{ BA}}\mathrm{e}^{+\mathrm{i}P(\boldsymbol{\lambda})x-\mathrm{i}E(\boldsymbol{\lambda})t}\frac{|\Phi(\boldsymbol{\lambda})\rangle\langle\Phi(\boldsymbol{\lambda})|}{\langle\Phi(\boldsymbol{\lambda})|\Phi(\boldsymbol{\lambda})\rangle}|\Phi_{0}(\boldsymbol{\mu})\rangle\nonumber \\
 & =\sum_{\boldsymbol{\lambda}:\text{ BA}}\mathrm{e}^{+\mathrm{i}P(\boldsymbol{\lambda})x-\mathrm{i}E(\boldsymbol{\lambda})t}\frac{\mathcal{M}(\boldsymbol{\lambda};\boldsymbol{\mu})}{\mathcal{N}(\boldsymbol{\lambda})}|\Phi(\boldsymbol{\lambda})\rangle\label{eq:int1}
\end{align}
where the sum runs over all solutions of the Bethe equations for $N$
particles in length $L$, and the eigenstate overlaps and norms are
\begin{align*}
\mathcal{M}(\boldsymbol{\lambda};\boldsymbol{\mu}) & \equiv\langle\Phi(\boldsymbol{\lambda})|\Phi_{0}(\boldsymbol{\mu})\rangle\\
\mathcal{N}(\boldsymbol{\lambda}) & \equiv\langle\Phi(\boldsymbol{\lambda})|\Phi(\boldsymbol{\lambda})\rangle
\end{align*}
From (\ref{eq:TG_wf}) it is straightforward to calculate the norm
$\mathcal{N}(\boldsymbol{\lambda})$
\begin{equation}
\mathcal{N}(\boldsymbol{\lambda})=L^{N}\label{eq:TG_norm}
\end{equation}

\subsection{Initial state overlaps \label{subsec:overlaps}}

The first problem we have to address is the calculation of the overlaps
$\mathcal{M}(\boldsymbol{\lambda};\boldsymbol{\mu})$.  From the
coordinate space form of the eigenstate wave-functions (\ref{eq:TG_wf})
we find 
\begin{align*}
\mathcal{M}(\boldsymbol{\lambda};\boldsymbol{\mu}) & =\int_{0}^{L/2}\mathrm{d}\boldsymbol{x}\;\langle\Phi(\boldsymbol{\lambda})|\boldsymbol{x}\rangle\langle\boldsymbol{x}|\Phi_{0}(\boldsymbol{\mu})\rangle\\
 & =\frac{1}{N!}\int_{0}^{L/2}\mathrm{d}\boldsymbol{x}\;\det_{i,j}\left[\exp\left(-\mathrm{i}\lambda_{j}x_{i}\right)\right]\det_{k,\ell}\left[\exp\left(\mathrm{i}\mu_{\ell}x_{k}\right)\right]\left(\prod_{j>i}\text{sign}(x_{j}-x_{i})\right)^{2}\\
 & =\frac{1}{N!}\int_{0}^{L/2}\mathrm{d}\boldsymbol{x}\;\det_{i,j}\left[\exp\left(-\mathrm{i}\lambda_{j}x_{i}\right)\right]\det_{k,\ell}\left[\exp\left(\mathrm{i}\mu_{\ell}x_{k}\right)\right]
\end{align*}
The last expression can be evaluated easily using the\emph{ Andr\'{e}ief
identity} \citep{Andreief}
\begin{equation}
\int\prod_{n=1}^{N}\mathrm{d}\mu(x_{n})\;\det_{i,j}\left[f_{j}\left(x_{i}\right)\right]\det_{k,\ell}\left[g_{\ell}\left(x_{k}\right)\right]=N!\det_{i,j}\left[\int\mathrm{d}\mu(x)\;f_{i}\left(x\right)g_{j}\left(x\right)\right]\label{eq:Andreief}
\end{equation}
which reduces the multiple integral of the product of two determinants
to the determinant of a matrix whose elements are single-variable
integrals. Using this identity we find
\begin{align}
\mathcal{M}(\boldsymbol{\lambda};\boldsymbol{\mu}) & =\det_{i,j}\left[\int_{0}^{L/2}\mathrm{d}x\;\exp\left(\mathrm{i}(\mu_{i}-\lambda_{j})x\right)\right]\label{eq:TGoverlaps2}
\end{align}
The application of the Andr\'{e}ief identity for the calculation of eigenstate
overlaps for this type of quench problems and the reduction to the
single-particle expression is a property that is valid for any effectively
non-interacting model as pointed out in \citep{DeLuca2015}, which
studied the same problem in the spin-$\nicefrac{1}{2}$ XX chain. 

Note that the integral appearing in the last expression is precisely
the same as in the overlap of the single-particle problem (\ref{eq:TGoverlaps})
\[
\int_{0}^{L/2}\mathrm{d}x\;\mathrm{e}^{\mathrm{i}(\mu-\lambda)x}=\frac{\mathrm{i}}{\lambda-\mu}\left({\rm e}^{{\rm i}(\mu-\lambda)L/2}-1\right)
\]
In order to evaluate the thermodynamic limit later on, we would like
to express the overlaps as continuous non-oscillatory functions of
the rapidities $\lambda_{i}$, which can be done by eliminating $L$
using the quantisation conditions. The quantised values of $\boldsymbol{\mu}$
and $\boldsymbol{\lambda}$ as dictated by (\ref{eq:TG_BA0}) and
(\ref{eq:TG_BA}) are not generally the same as in the single-particle
problem. If $N$ is odd then $\mu_{i}$ and $\lambda_{j}$ are integer
multiples of $2\pi/(L/2)$ and $2\pi/L$ respectively, exactly as
before (\ref{eq:TGoverlaps}). If however $N$ is even then they are
half-integer multiples, which means that they can never be equal to
each other (the difference of two integers cannot be $\tfrac{1}{2}$).
Let us focus on this case of even $N$. We find
\begin{align}
\int_{0}^{L/2}\mathrm{d}x\;\mathrm{e}^{\mathrm{i}(\mu-\lambda)x} & =\frac{1}{\lambda-\mu}\times\begin{cases}
-1-\mathrm{i} & \text{ if }n=\frac{\lambda L}{2\pi}-\tfrac{1}{2}\text{ is even}\\
+1-\mathrm{i} & \text{ if }n\text{ is odd}
\end{cases}\label{eq:TGoverlaps3}\\
 & =-\frac{\mathrm{i}+s}{\lambda-\mu},\qquad\text{with }s=\pm1\text{ for }n\text{ even/odd.}\nonumber 
\end{align}
We therefore need again to introduce an index $s_{j}$ to distinguish
between even and odd sectors ($s_{j}=\pm1$) separately for each rapidity
variable $\lambda_{j}$ and redefine the overlaps as functions of
both the rapidity vectors $\boldsymbol{\lambda},\boldsymbol{\mu}$
and an additional vector of discrete indices $\boldsymbol{s}$ with
each of the $s_{j}$ taking one of the two values depending on $\lambda_{j}$.
We finally find that, for even $N$ 
\begin{align}
\mathcal{M}_{\boldsymbol{s}}(\boldsymbol{\lambda};\boldsymbol{\mu}) & =\det_{i,j}\left[-\frac{\mathrm{i}+s_{j}}{\lambda_{j}-\mu_{i}}\right]\label{eq:TGoverlaps3b}
\end{align}
Considering the variables $\boldsymbol{\lambda}$ as continuous, we
observe that, analogously to the single-particle case, the overlaps
$\mathcal{M}_{\boldsymbol{s}}(\boldsymbol{\lambda};\boldsymbol{\mu})$
have simple poles when any of the rapidities $\lambda_{i}$ tends
to any of the $\mu_{j}$ i.e. they exhibit $N$-dimensional poles
at $\boldsymbol{\lambda}\to\boldsymbol{\mu}$ and at all permutations
of components of $\boldsymbol{\mu}$. As in the single particle calculation,
these poles are of ``kinematical'' type and reflect the fact that
in the thermodynamic limit the complete set of particle momenta is
conserved.

\subsection{Eigenstate summation through multivariable version of Cauchy's integral
formula \label{subsec:Cauchy}}

The next step is to perform the summation over post-quench eigenstates.
This can be done again rigorously using the same contour integral
trick as in the single particle case, the only difference being that
we now need a multivariable version of it. For odd $N$, the rapidities
$\lambda_{j}$ are integer multiples of $2\pi/L$ as in the previously
studied single particle case, and we have already seen that the even
sector should be treated differently from the the odd sector, due
to the exceptional property that in the even sector the overlaps are
simply proportional to a Kronecker delta. Let us therefore focus now
on the case of even $N$, which is different. In this case we still
have to disinguish between the two sectors, but the even sector is
not exceptional and can be treated in the same way as the odd. As
before, to pick the values of rapidities allowed by the quantisation
conditions (\ref{eq:TG_BA}), we introduce a suitable function $F_{\boldsymbol{s}}(\boldsymbol{\lambda})$
that has simple poles of residue $1/(2\pi\mathrm{i})$ when each of
the $\lambda_{j}$ is a half-integer multiple of $2\pi/L$, and in
addition distinguishes between odd and even integers through the corresponding
index $s_{j}$. More specifically, we choose
\begin{align}
F_{\boldsymbol{s}}(\boldsymbol{\lambda}) & \equiv\left(-\frac{L}{4\pi}\right)^{N}\prod_{i=1}^{N}\frac{1}{\mathrm{i}s_{i}\mathrm{e}^{\mathrm{i}\lambda_{i}L/2}+1}\qquad\text{for even }N\label{eq:FNeven}
\end{align}
We can easily verify that (\ref{eq:FNeven}) considered as a function
of $\lambda_{j}$ has poles at $2\pi/L(n_{j}+\tfrac{1}{2})$ with
$n_{j}$ even or odd for $s_{j}=\pm1$ respectively. 

We can now use a multivariable version of the residue theorem to express
the multiple summation over allowed rapidities, as an integral over
a continuous function of the $\boldsymbol{\lambda}$'s. More specifically,
given a function $F(\boldsymbol{\lambda})=1/\prod_{i=1}^{N}f_{i}(\boldsymbol{\lambda})$
such that each of the $f_{i}(\boldsymbol{\lambda})$ has a simple
zero at a single point $\boldsymbol{\lambda}=\boldsymbol{\lambda}^{*}$,
then for any multi-dimensional contour $\boldsymbol{C}=C_{1}\times C_{2}\times\dots\times C_{N}$
encircling this point and any function $g(\boldsymbol{\lambda})$
of the rapidities $\boldsymbol{\lambda}$ that is analytic inside
the contour $\boldsymbol{C}$, we have (see e.g. appendix A of \citep{Pozsgay-Takacs})
\begin{equation}
\oint_{\boldsymbol{C}}\frac{{\rm d}^{N}\boldsymbol{\lambda}}{(2\pi{\rm i})^{N}}\,F(\boldsymbol{\lambda})g(\boldsymbol{\lambda})=\frac{g(\boldsymbol{\lambda}^{*})}{\left.\det\left(\frac{\partial f_{i}}{\partial\lambda_{j}}\right)\right|_{\boldsymbol{\lambda}=\boldsymbol{\lambda}^{*}}}\label{eq:trick}
\end{equation}
This can be generalised to the case of more than one zeroes, i.e.
when the functions $f_{i}(\boldsymbol{\lambda})$ have simple zeroes
at each one of a set of points $\{\boldsymbol{\lambda}_{\alpha}^{*}\}$,
in which case the right hand side of the above equation should be
replaced by a sum over all these points with appropriate signs, determined
by the direction of integration resulting from the contour deformation.
If on the other hand the function $g(\boldsymbol{\lambda})$ has
additional poles inside the contour $\boldsymbol{C}$, not coinciding
with any of the points $\{\boldsymbol{\lambda}_{\alpha}^{*}\}$, then
the contours in the above formula should be modified to exclude these
additional poles, or equivalently their residues must be subtracted
off. 

Using this method, the sum over post-quench eigenstates can be written
in integral form as follows
\[
\sum_{\boldsymbol{\lambda}:\text{ BA}}\dots\,=\frac{1}{N!}\sum_{\boldsymbol{s}}\oint_{\boldsymbol{C}}{\rm d}^{N}\boldsymbol{\lambda}\,F_{\boldsymbol{s}}(\boldsymbol{\lambda})\dots
\]
where the contours $C_{i}$ in each of the above integrals enclose
the real rapidity axis where all allowed rapidities lie, i.e. should
be chosen to consist of two straight lines one just above and one
just below the real axis. However, because the overlaps $\mathcal{M}_{\boldsymbol{s}}(\boldsymbol{\lambda};\boldsymbol{\mu})$
have multi-dimensional poles at $\boldsymbol{\lambda}\to\boldsymbol{\mu}$
and permutations, these points must be excluded by subtracting from
each contour small circles around them or equivalently by subtracting
the corresponding residues. The $1/N!$ factor accounts for permutations
of the order of the rapidities $\boldsymbol{\lambda}$ which correspond
to identical Bethe states, and the sum over the discrete indices $\boldsymbol{s}$
accounts for all combinations of the even and odd sectors for each
of the rapidities. 

We can therefore write (\ref{eq:int1}) as 
\begin{align}
\mathrm{e}^{+\mathrm{i}Px-\mathrm{i}Ht}|\Phi_{0}(\boldsymbol{\mu})\rangle & =\sum_{\boldsymbol{\lambda}:\text{ BA}}\mathrm{e}^{+\mathrm{i}P(\boldsymbol{\lambda})x-\mathrm{i}E(\boldsymbol{\lambda})t}\frac{\mathcal{M}(\boldsymbol{\lambda};\boldsymbol{\mu})}{\mathcal{N}(\boldsymbol{\lambda})}|\Phi(\boldsymbol{\lambda})\rangle\nonumber \\
 & =\frac{1}{N!}\sum_{\boldsymbol{s}}\oint_{\boldsymbol{C}}{\rm d}^{N}\boldsymbol{\lambda}\,F_{\boldsymbol{s}}(\boldsymbol{\lambda})\mathrm{e}^{+\mathrm{i}P(\boldsymbol{\lambda})x-\mathrm{i}E(\boldsymbol{\lambda})t}\frac{\mathcal{M}_{\boldsymbol{s}}(\boldsymbol{\lambda};\boldsymbol{\mu})}{\mathcal{N}(\boldsymbol{\lambda})}|\Phi(\boldsymbol{\lambda})\rangle\label{eq:int2}
\end{align}
Note that the poles at $\boldsymbol{\lambda}\to\boldsymbol{\mu}$
and permutations are between those of $F_{\boldsymbol{s}}(\boldsymbol{\lambda})$
i.e. the rapidity combinations allowed by the full system's Bethe
equations, and do not coincide with any of them for even $N$. This
is because, as mentioned above, the allowed rapidities $\boldsymbol{\lambda}$
and $\boldsymbol{\mu}$ are half-integer multiples of $2\pi/L$ and
$2\pi/(L/2)$ respectively, so they can never be equal to each other.
Also note that all other constituents of the integrand in (\ref{eq:int2})
i.e. the function $\mathrm{e}^{+\mathrm{i}P(\boldsymbol{\lambda})x-\mathrm{i}E(\boldsymbol{\lambda})t}$
and the state $|\Phi(\boldsymbol{\lambda})\rangle$ itself, are analytic
functions for real rapidities and do not introduce any other singularities.
Lastly, the dependence on the discrete multi-variable $\boldsymbol{s}$
is limited to $F_{\boldsymbol{s}}(\boldsymbol{\lambda})$ and $\mathcal{M}_{\boldsymbol{s}}(\boldsymbol{\lambda};\boldsymbol{\mu})$
only.

At this step we can also evaluate the sum over the discrete indices
$\boldsymbol{s}$. By first absorbing the product in (\ref{eq:FNeven})
and the sum over $\boldsymbol{s}$ into the determinant (\ref{eq:TGoverlaps3b}),
using the properties 
\[
\prod_{i=1}^{N}f_{i}\det\left(A_{ij}\right)=\det\left(f_{i}A_{ij}\right)
\]
and
\[
\sum_{\boldsymbol{s}}\det\left(A_{ij}(s_{j})\right)=\det\left(\sum_{s}A_{ij}(s)\right)
\]
we obtain \foreignlanguage{english}{
\begin{align*}
\sum_{\boldsymbol{s}}F_{\boldsymbol{s}}(\boldsymbol{\lambda})M_{\boldsymbol{s}}(\boldsymbol{\lambda};\boldsymbol{\mu}) & =\sum_{\boldsymbol{s}}\left(-\frac{L}{4\pi}\right)^{N}\prod_{j=1}^{N}\frac{1}{\mathrm{i}s_{j}\mathrm{e}^{\mathrm{i}\lambda_{j}L/2}+1}\det_{i,j}\left[-\frac{\mathrm{i}+s_{j}}{\lambda_{j}-\mu_{i}}\right]\\
 & =\left(\frac{L}{4\pi}\right)^{N}\det_{i,j}\left[\sum_{s}\frac{1}{\mathrm{i}s\mathrm{e}^{\mathrm{i}\lambda_{j}L/2}+1}\left(\frac{\mathrm{i}+s}{\lambda_{j}-\mu_{i}}\right)\right]\\
 & =\left(\frac{L}{2\pi}\right)^{N}\det_{i,j}\left[\frac{1-\mathrm{e}^{+\mathrm{i}\lambda_{j}L/2}}{1+\mathrm{e}^{+\mathrm{i}\lambda_{j}L}}\left(\frac{{\rm i}}{\lambda_{j}-\mu_{i}}\right)\right]
\end{align*}
and after substitution into (\ref{eq:int2}) }
\begin{align}
\mathrm{e}^{+\mathrm{i}Px-\mathrm{i}Ht}|\Phi_{0}(\boldsymbol{\mu})\rangle & =\frac{1}{N!}\oint_{\boldsymbol{C}}\frac{{\rm d}^{N}\boldsymbol{\lambda}}{(2\pi)^{N}}\,\mathrm{e}^{+\mathrm{i}P(\boldsymbol{\lambda})x-\mathrm{i}E(\boldsymbol{\lambda})t}\prod_{j=1}^{N}\frac{1-\mathrm{e}^{+\mathrm{i}\lambda_{j}L/2}}{1+\mathrm{e}^{+\mathrm{i}\lambda_{j}L}}\det_{i,j}\left[\frac{{\rm i}}{\lambda_{j}-\mu_{i}}\right]|\Phi(\boldsymbol{\lambda})\rangle\label{eq:int2b}
\end{align}
Note that in the last formula the function $F_{\boldsymbol{s}}(\boldsymbol{\lambda})$
appears to be replaced by a function with poles at the eigenvalues
of both the even and odd sector eigenstates, which might suggest that
the splitting into these two sectors is not that relevant in the end.
However, the expression that we would obtain if we followed all earlier
steps except for the splitting would be different and incorrect, even
though still characterised by the same pole structure. 

\subsection{Multivariable Kinematical Pole Residue \label{subsec:pole}}

Before we analyse the asymptotics of our contour integral formula
in the thermodynamic and large space and time limit, let us focus
on the multi-dimensional kinematical pole at $\boldsymbol{\lambda}\to\boldsymbol{\mu}$
and calculate its residue. As anticipated based on the single particle
calculation, this pole gives the only non-vanishing contribution of
the contour integral in the above double limit. 

Starting from (\ref{eq:TGoverlaps3b}), the residue of the overlap
$\mathcal{M}_{\boldsymbol{s}}(\boldsymbol{\lambda};\boldsymbol{\mu})$
at $\boldsymbol{\lambda}\to\boldsymbol{\mu}$ (and at any permutation
$\boldsymbol{\mu}_{\pi}$ of $\boldsymbol{\mu}$) can be easily evaluated
by expanding the determinant as a sum over permutations and noticing
that exactly one of them and only that contributes a non-vanishing
residue. Explicitly, we obtain
\begin{align}
\underset{\boldsymbol{\lambda}=\boldsymbol{\mu}}{\text{Res}}\mathcal{M}_{\boldsymbol{s}}(\boldsymbol{\lambda};\boldsymbol{\mu}) & =\sum_{\text{all perm. }\pi}(-1)^{[\pi]}\underset{\boldsymbol{\lambda}=\boldsymbol{\mu}}{\text{Res}}\prod_{j=1}^{N}\left(-\frac{\mathrm{i}+s_{j}}{\lambda_{j}-\mu_{\pi{}_{i}}}\right)\nonumber \\
 & =(-1)^{N}\prod_{j=1}^{N}(\mathrm{i}+s_{j})\label{eq:pole1}
\end{align}
The residue at $\boldsymbol{\lambda}=\boldsymbol{\mu}_{\pi}$ is the
same for all permutations $\pi$, except for a sign which is equal
to the signature of the permutation. 

Moreover the value of the function $F_{\boldsymbol{s}}(\boldsymbol{\lambda})$
at $\boldsymbol{\lambda}=\boldsymbol{\mu}$ and permutations can be
obtained taking into account that the rapidities $\boldsymbol{\mu}$ satisfy
the half system Bethe equations (\ref{eq:TG_BA0}) for even $N$ 
\begin{align}
F_{\boldsymbol{s}}(\boldsymbol{\mu}_{\pi}) & =\left(-\frac{L}{4\pi}\right)^{N}\prod_{i=1}^{N}\frac{1}{\mathrm{i}s_{i}\mathrm{e}^{\mathrm{i}\mu_{\pi_{i}}L/2}+1}\nonumber \\
 & =\left(-\frac{L}{4\pi}\right)^{N}\prod_{i=1}^{N}\frac{1}{\mathrm{i}s_{i}(-1)^{N-1}+1}\nonumber \\
 & =\left(-\frac{L}{4\pi}\right)^{N}\prod_{i=1}^{N}\frac{1}{-\mathrm{i}s_{i}+1}\label{eq:pole2}
\end{align}
Notice that there is no oscillating phase left in
the last expression. By combining (\ref{eq:pole1}) and (\ref{eq:pole2})
we obtain
\begin{align}
\underset{\boldsymbol{\lambda}=\boldsymbol{\mu}}{\text{Res}}F_{\boldsymbol{s}}(\boldsymbol{\mu})\mathcal{M}_{\boldsymbol{s}}(\boldsymbol{\lambda};\boldsymbol{\mu}) & =\left(\frac{L}{4\pi}\right)^{N}\prod_{i=1}^{N}\frac{\mathrm{i}+s_{i}}{-\mathrm{i}s_{i}+1}\nonumber \\
 & =\mathrm{i}^{N}\left(\frac{L}{4\pi}\right)^{N}\label{eq:pole3}
\end{align}
Notice that this expression is independent of $\boldsymbol{s}$, therefore
the corresponding discrete index sum becomes trivial. From the above
we conclude that the contribution of the kinematical poles to the
multiple contour integral (\ref{eq:int2}) is given by the simple
result
\[
\frac{1}{N!}\sum_{\boldsymbol{s}}\underset{\boldsymbol{\lambda}=\boldsymbol{\mu}}{\text{Res}}F_{\boldsymbol{s}}(\boldsymbol{\lambda})\mathrm{e}^{+\mathrm{i}P(\boldsymbol{\lambda})x-\mathrm{i}E(\boldsymbol{\lambda})t}\frac{\mathcal{M}_{\boldsymbol{s}}(\boldsymbol{\lambda};\boldsymbol{\mu})}{\mathcal{N}(\boldsymbol{\lambda})}|\Phi(\boldsymbol{\lambda})\rangle=\mathrm{e}^{+\mathrm{i}P(\boldsymbol{\mu})x-\mathrm{i}E(\boldsymbol{\mu})t}|\Phi(\boldsymbol{\mu})\rangle
\]
where the $N!$ multiplicity of the poles has been taken into account.

\subsection{Asymptotics of the integral in the thermodynamic limit \label{subsec:TDL}}

We are now ready to calculate the asymptotic form of the contour integral
representation (\ref{eq:int2}) in the thermodynamic limit. As expected,
this step will result in an $N$-dimensional line integral in the
positive direction, with a specific prescription on how to pass around
the kinematical poles. As we have seen in the single particle case,
this prescription is going to be crucial at the next step, which is
the derivation of the asymptotics in the large $x,t$ limit.

Starting from (\ref{eq:int2}), we split each of the contours composing
the multi-dimensional contour in $\boldsymbol{C}$ into their three
parts: the straight line $\mathbb{R}-{\rm i}\epsilon$ below the real
axis in the positive direction ($C_{-}$), the straight line $\mathbb{R}+{\rm i}\epsilon$
above the real axis in the negative direction ($C_{+}$) and the circles
$C(\mu_{i},\epsilon)$ centred at each of the points $\mu_{i}$ with
a small radius $\epsilon$ in the negative direction (collectively
denoted as $C_{p}$), where in all cases $\epsilon>0$. With this
decomposition of the contours, expanding the integral we obtain a
sum of $3^{N}$ cross-terms, that is the sum of all possible combinations
of integrals of the $N$ rapidity variables along the three different
parts of the contour. We will now show that in the thermodynamic limit,
only $2^{N}$ terms survive, specifically those of the above terms
that do not involve any integral along the first contour part $C_{-}=\mathbb{R}-{\rm i}\epsilon$.
To see why this is true, we simply have to examine the behaviour of
the function $F_{\boldsymbol{s}}(\boldsymbol{\lambda})$ when at least
one of the rapidities lies along $C_{-}$ in the limit $L\to\infty$:
we find that this limit results in an exponentially decaying factor
that suppresses all such terms, while all others remain finite and
even simplify.

Indeed, let us consider $F_{\boldsymbol{s}}(\boldsymbol{\lambda})$
given by (\ref{eq:FNeven}) as a function of one of the rapidities,
$\lambda_{i}$. Firstly, we notice that
\[
\lim_{L\to\infty}\frac{1}{\mathrm{i}s_{i}\mathrm{e}^{\mathrm{i}\lambda_{i}L/2}+1}=\begin{cases}
1 & \text{ if Im}(\lambda_{i})>0\\
0 & \text{ if Im}(\lambda_{i})<0
\end{cases}
\]
independently of the value of $s_{i}$. Actually it is much more useful
to analyse the asymptotics towards these limits, which are exponentially
decaying functions of $L$ in both cases
\[
\frac{1}{\mathrm{i}s_{i}\mathrm{e}^{\mathrm{i}\lambda_{i}L/2}+1}\sim_{L\to\infty}\begin{cases}
1+\mathcal{O}(\mathrm{e}^{-\epsilon L/2}) & \text{ if }\lambda_{i}\in\mathbb{R}+{\rm i}\epsilon\\
\mathcal{O}(\mathrm{e}^{-\epsilon L/2}) & \text{ if }\lambda_{i}\in\mathbb{R}-{\rm i}\epsilon
\end{cases}\,
\]
If $\lambda_{i}\in\mathbb{R}$ the limit is generally indeterminate
and depends on the value of $\lambda_{i}$, but when it is equal to
one of the $\mu_{i}$'s we have already seen in (\ref{eq:pole2})
that it tends to the finite value $1/(-\mathrm{i}s_{i}+1)$ independently
of $L$. From these observations, we immediately see that in the limit
$L\to\infty$ the function $F_{\boldsymbol{s}}(\boldsymbol{\lambda})$
has a finite value plus $\mathcal{O}(\mathrm{e}^{-\epsilon L/2})$
corrections if all of the rapidities $\boldsymbol{\lambda}$ are either
on $C_{-}$ or at one of the points $\mu_{j}$ (i.e. at the kinematical
poles), while it decays as $\mathcal{O}(\mathrm{e}^{-\epsilon nL/2})$
if $n$ out of the $N$ rapidities $\boldsymbol{\lambda}$ are on
$C_{+}$. Moreover, the surviving $2^{N}$ terms can be recast in
the simpler form
\begin{equation}
\mathrm{e}^{+\mathrm{i}Px-\mathrm{i}Ht}|\Phi_{0}(\boldsymbol{\mu})\rangle=\frac{1}{N!}\left(\frac{L}{2}\right)^{N}\sum_{\boldsymbol{s}}\prod_{i=1}^{N}\left(\intop_{-\infty+\mathrm{i}\epsilon}^{+\infty+\mathrm{i}\epsilon}\frac{{\rm d}\lambda_{i}}{2\pi}+\frac{1}{1-\mathrm{i}s_{i}}\oint_{C_{p}}\frac{{\rm d}\lambda_{i}}{2\pi}\right)\,\mathrm{e}^{+\mathrm{i}P(\boldsymbol{\lambda})x-\mathrm{i}E(\boldsymbol{\lambda})t}\frac{\mathcal{M}_{\boldsymbol{s}}(\boldsymbol{\lambda};\boldsymbol{\mu})}{\mathcal{N}(\boldsymbol{\lambda})}|\Phi(\boldsymbol{\lambda})\rangle+\mathcal{O}(\mathrm{e}^{-\epsilon L/2})\label{eq:TG_TDL1}
\end{equation}
Note that we have used the minus signs $(-1)^{N}$ to reverse the
integration direction in both $C_{+}$ and $C_{p}$, so that we now
have line integrals along $\mathbb{R}+{\rm i}\epsilon$ running in
the positive direction and circular integrals around the poles also
in the positive direction. 

Obviously, we would arrive at an equivalent but more explicit formula
if we started from the simplified version (\ref{eq:int2b}) and applied
the same analysis. Explicitly, using the asymptotics
\[
\frac{1-\mathrm{e}^{+\mathrm{i}\lambda_{j}L/2}}{1+\mathrm{e}^{+\mathrm{i}\lambda_{j}L}}\sim_{L\to\infty}\begin{cases}
1+\mathcal{O}(\mathrm{e}^{-\epsilon L/2}) & \text{ if }\lambda_{i}\in\mathbb{R}+{\rm i}\epsilon\\
\mathcal{O}(\mathrm{e}^{-\epsilon L/2}) & \text{ if }\lambda_{i}\in\mathbb{R}-{\rm i}\epsilon
\end{cases}\,
\]
we obtain the final result
\begin{equation}
\mathrm{e}^{+\mathrm{i}Px-\mathrm{i}Ht}|\Phi_{0}(\boldsymbol{\mu})\rangle=\frac{1}{N!}\int_{\boldsymbol{C}'}\frac{{\rm d}^{N}\boldsymbol{\lambda}}{(2\pi)^{N}}\,\mathrm{e}^{+\mathrm{i}P(\boldsymbol{\lambda})x-\mathrm{i}E(\boldsymbol{\lambda})t}\det_{i,j}\left[\frac{{\rm i}}{\lambda_{j}-\mu_{i}}\right]|\Phi(\boldsymbol{\lambda})\rangle+\mathcal{O}(\mathrm{e}^{-\epsilon L/2})\label{eq:TG_TDL2}
\end{equation}
where for brevity we denote the integration measure as 
\[
\int_{\boldsymbol{C}'}\frac{{\rm d}^{N}\boldsymbol{\lambda}}{(2\pi)^{N}}\equiv\prod_{i=1}^{N}\left(\intop_{-\infty+\mathrm{i}\epsilon}^{+\infty+\mathrm{i}\epsilon}+\oint_{C_{p}}\right)\frac{{\rm d}\lambda_{i}}{2\pi}
\]
and used also that $(-1)^{N}=1$ for even $N$.

It is important that we do not simply derive the limit $L\to\infty$,
but also the exact asymptotics, i.e. the exponential bound on corrections
to this limit. This is crucial for taking the thermodynamic limit
$N,L\to\infty$ at fixed density $n=N/L$ as it allows us to show
that the two sides of (\ref{eq:TG_TDL2}) have the same thermodynamic
limit, even without ever taking $N\to\infty$ explicitly, which would
result in the multiple rapidity integral turning into the less practical
form of a functional integral. In other words, we have shown that
the thermodynamic limit of $\mathrm{e}^{+\mathrm{i}Px-\mathrm{i}Ht}|\Phi_{0}(\boldsymbol{\mu})\rangle$
reduces to 
\begin{equation}
\lim_{{N,L\to\infty\atop n:\text{ fixed}}}\mathrm{e}^{+\mathrm{i}Px-\mathrm{i}Ht}|\Phi_{0}(\boldsymbol{\mu})\rangle=\lim_{N\to\infty}\frac{1}{N!}\int_{\boldsymbol{C}'}\frac{{\rm d}^{N}\boldsymbol{\lambda}}{(2\pi)^{N}}\,\mathrm{e}^{+\mathrm{i}P(\boldsymbol{\lambda})x-\mathrm{i}E(\boldsymbol{\lambda})t}\det_{i,j}\left[\frac{{\rm i}}{\lambda_{j}-\mu_{i}}\right]|\Phi(\boldsymbol{\lambda})\rangle\label{eq:TG_TDL3}
\end{equation}
which is one of our main results. Not only is this limit simpler than
the original, but also the functional form of the integrand has been
greatly simplified. In particular the awkward dependence on $L$ of
the original expression, which was most responsible for the highly
oscillatory behaviour of the integrand, has been completely eliminated.
Lastly let us remark that, as can be easily verified, the multi-dimensional
residues at the kinematical poles are equal to the values derived
in the previous step.

\subsection{Asymptotics at large distances \& times \label{subsec:asympt}}

Our last step is the derivation of the asymptotics at large distances
and times. At this point it is convenient to apply the projection
to the local basis $\langle\boldsymbol{z}|$, which has been implicitly
assumed from the beginning, and use the coordinate space wavefunction
of the Bethe states (\ref{eq:TG_wf}). The last result (\ref{eq:TG_TDL3})
now reads
\begin{align*}
\lim_{{N,L\to\infty\atop n:\text{ fixed}}}\langle\boldsymbol{z}|\mathrm{e}^{+\mathrm{i}Px-\mathrm{i}Ht}|\Phi_{0}(\boldsymbol{\mu})\rangle & =\lim_{N\to\infty}\frac{1}{N!}\int_{\boldsymbol{C}'}\frac{{\rm d}^{N}\boldsymbol{\lambda}}{(2\pi)^{N}}\,\mathrm{e}^{+\mathrm{i}P(\boldsymbol{\lambda})x-\mathrm{i}E(\boldsymbol{\lambda})t}\det_{i,j}\left[\frac{{\rm i}}{\lambda_{j}-\mu_{i}}\right]\langle\boldsymbol{z}|\Phi(\boldsymbol{\lambda})\rangle\\
 & =\lim_{N\to\infty}\frac{1}{N!}\prod_{j>i}\text{sign}(z_{j}-z_{i})\int_{\boldsymbol{C}'}\frac{{\rm d}^{N}\boldsymbol{\lambda}}{(2\pi)^{N}}\,\mathrm{e}^{+\mathrm{i}P(\boldsymbol{\lambda})x-\mathrm{i}E(\boldsymbol{\lambda})t}\det_{i,j}\left[\frac{{\rm i}}{\lambda_{j}-\mu_{i}}\right]\det_{k,l}\left[\exp\left(\mathrm{i}\lambda_{k}z_{l}\right)\right]
\end{align*}
Of course we can immediately see that, using the Andr\'{e}ief identity
(\ref{eq:Andreief}), the last expression gives
\begin{align*}
\lim_{{N,L\to\infty\atop n:\text{ fixed}}}\langle\boldsymbol{z}|\mathrm{e}^{+\mathrm{i}Px-\mathrm{i}Ht}|\Phi_{0}(\boldsymbol{\mu})\rangle & =\lim_{N\to\infty}\prod_{j>i}\text{sign}(z_{j}-z_{i})\det_{i,j}\int_{C'}\frac{{\rm d}\lambda}{2\pi}\,\mathrm{e}^{+\mathrm{i}p(\lambda)x-\mathrm{i}e(\lambda)t}\frac{{\rm i}}{\lambda-\mu_{i}}\mathrm{e}^{\mathrm{i}\lambda z_{j}}
\end{align*}
which reduces the problem to precisely the single particle problem.
Therefore the asymptotic analysis is exactly the same as before, giving
\begin{align}
\lim_{{x,t\to\infty\atop x/t=\xi:\text{ fixed}}}\lim_{{N,L\to\infty\atop n:\text{ fixed}}}\mathrm{e}^{-\mathrm{i}P(\boldsymbol{\mu})x+\mathrm{i}E(\boldsymbol{\mu})t}\langle\boldsymbol{z}|\mathrm{e}^{+\mathrm{i}Px-\mathrm{i}Ht}|\Phi_{0}(\boldsymbol{\mu})\rangle & =\lim_{N\to\infty}\prod_{i=1}^{N}\Theta(\xi-v(\mu_{i}))\langle\boldsymbol{z}|\Phi(\boldsymbol{\mu})\rangle\label{eq:final_res}
\end{align}

Alternatively, if we do not want to rely on the Andr\'{e}ief identity,
we can repeat the steps of the asymptotic analysis for the multivariate
case. First of all, considering each integral separately it is clear
that the real part of the exponent of the exponential function switches
sign at the point $\lambda_{*}(\xi)$ that is the solution of the
equation
\begin{equation}
v(\lambda_{*})=\xi\label{eq:threshold}
\end{equation}
where 
\[
v(\lambda)\equiv\frac{E'(\lambda)}{P'(\lambda)}
\]
is the group velocity. Therefore in order to handle the $x,t\to\infty$
limit, we deform the integration contours so that they always lie
in the region where the real part of the exponent is negative, taking
into account the contribution of the kinematical pole if it is eventually
crossed due to the deformation. Explicitly, the contour of the line
integral becomes
\[
\intop_{-\infty+\mathrm{i}\epsilon}^{+\infty+\mathrm{i}\epsilon}\frac{{\rm d}\lambda}{2\pi}=\left(\intop_{-\infty+\mathrm{i}\epsilon}^{\lambda_{*}(\xi)+\mathrm{i}\epsilon}+\intop_{\lambda_{*}(\xi)+\mathrm{i}\epsilon}^{\lambda_{*}(\xi)-\mathrm{i}\epsilon}+\intop_{\lambda_{*}(\xi)-\mathrm{i}\epsilon}^{+\infty-\mathrm{i}\epsilon}\right)\frac{{\rm d}\lambda}{2\pi}-\oint_{C_{p}}\frac{{\rm d}\lambda}{2\pi}\,\Theta(v(\lambda)-\xi)
\]
and combined with the earlier contribution of the circular integral
around the pole 
\[
\left(\intop_{-\infty+\mathrm{i}\epsilon}^{+\infty+\mathrm{i}\epsilon}+\oint_{C_{p}}\right)\frac{{\rm d}\lambda}{2\pi}=\left(\intop_{-\infty+\mathrm{i}\epsilon}^{\lambda_{*}(\xi)+\mathrm{i}\epsilon}+\intop_{\lambda_{*}(\xi)+\mathrm{i}\epsilon}^{\lambda_{*}(\xi)-\mathrm{i}\epsilon}+\intop_{\lambda_{*}(\xi)-\mathrm{i}\epsilon}^{+\infty-\mathrm{i}\epsilon}\right)\frac{{\rm d}\lambda}{2\pi}+\oint_{C_{p}}\frac{{\rm d}\lambda}{2\pi}\,\Theta(\xi-v(\lambda))
\]
We therefore split the contour $C'$ into two parts: the stair-like
contour $C''_{\lambda_{*}(\xi)}\equiv(-\infty+\mathrm{i}\epsilon,\lambda_{*}(\xi)+\mathrm{i}\epsilon]\cup[\lambda_{*}(\xi)+\mathrm{i}\epsilon,\lambda_{*}(\xi)-\mathrm{i}\epsilon]\cup[\lambda_{*}(\xi)-\mathrm{i}\epsilon,+\infty-\mathrm{i}\epsilon,)$
and the usual circular contour around the pole $C_{p}$ now coming
with a conditional measure $\Theta(\lambda_{*}(\xi)-\lambda)=\Theta(\xi-v(\lambda))$.
With this decomposition of the contours, expanding the multiple integral
we obtain a sum of $2^{N}$ cross-terms, that are all possible combinations
of integrals of the $N$ rapidity variables along the two different
parts of the contour. If the threshold value $\lambda_{*}(\xi)$ is
chosen as the solution of (\ref{eq:threshold}), then all terms involving
at least one integral along $C''_{\lambda_{*}(\xi)}$ decay in the
large distance and time limit, since the integrand decays exponentially
(line integrals parallel to the real axis) or at most algebraically
(short line integral perpendicular to the real axis). Therefore the
only surviving contribution comes from the cross-term that contains
only circular integrals around the kinematical pole as factors, which
is equal to the corresponding pole residue calculated earlier. In
this way we find again our final result (\ref{eq:final_res}).

Obviously it is only choosing the appropriate value for $\lambda_{*}$
that guarantees that the asymptotics in the combined limit is given
by the single cross-term corresponding to the multidimensional kinematical
pole contribution. If $\lambda_{*}$ is chosen correctly then the
next to leading order corrections must decay. As in the previous section
where we evaluated the thermodynamic limit, these corrections come
from all cross-terms between $(N-1)$ factors corresponding to pole
residues and one factor corresponding to a line integral along the
stair-shaped contour $C''_{\lambda_{*}(\xi)}$
\[
\sum_{j=1}^{N}\mathrm{e}^{+\mathrm{i}\sum_{i(\neq j)}p(\mu_{i})x-e(\mu_{i})t}\prod_{i(\neq j)}\Theta(\lambda_{*}(\xi)-\mu_{i})\intop_{C''_{\lambda_{*}(\xi)}}\frac{{\rm d}\lambda}{2\pi}\,\mathrm{e}^{+\mathrm{i}p(\lambda)x-\mathrm{i}e(\lambda)t}\langle\boldsymbol{z}|\Phi(\mu_{1},\dots,\mu_{i-1},\lambda,\mu_{i+1},\dots,\mu_{N})\rangle
\]
Therefore the condition for $\lambda_{*}$ can be phrased differently
as the condition for which the above expression vanishes in the combined
limit. The $\lambda$-independent coefficients of this sum that appear
before the integral are oscillatory but non-decaying. Since the coordinate
space wavefunction $\langle\boldsymbol{z}|\Phi\rangle$ essentially
is a determinant, i.e. a sum of products of factors of which only
one is $\lambda$ dependent, for each of these terms the integral
can be performed independently from the others terms. As a consequence,
the asymptotics of the integral is the same for each of these terms
and the same as above, therefore the threshold value is indeed given
by (\ref{eq:threshold}).

Eq. (\ref{eq:final_res}) is our final result for the asymptotics
of $\mathcal{K}(\boldsymbol{\mu};\boldsymbol{z};x,t)$ in the combined
limit, which as explained in sec. \ref{sec:Model} establishes the
emergence of the NESS and provides a complete characterisation of
it in terms of initial state information.

\section{Discussion\label{sec:Discussion}}

In this work we have developed a new analytical method for the derivation
of asymptotics of local observables after a geometric quench, a special
but representative type of inhomogenerous quench, focusing for the
moment on the non-interacting case. From the detailed calculation
presented here, we can draw certain conclusions that are instructive
on how to proceed to the more interesting and highly non-trivial interacting
case.

First of all, the derivation of the asymptotics does not rely significantly
on special characteristics of the initial state or the observable
considered. The only necessary information about these that we used
is that the initial state is diagonal in the pre-quench basis and
the observables are local. This suggests that our strategy of considering
the asymptotics of each pre-quench eigenstate after time evolution
under the post-quench Hamiltonian and projection onto the local basis
should be suitable for the interacting case as well.

Our method was mainly based on analyticity properties of the pre-
and post-quench eigenstate overlaps, especially on the kinematical
poles at equal momenta. Their presence is a direct consequence of
the step-like form of the initial state inhomogeneity as reflected
in momentum space and it is a characteristic feature of any inhomogeneous
quench of this type (including ``domain wall'' states and smooth
step-like initial density or temperature profiles). The necessity
for decomposition of the post-quench eigenstates into parity sectors
is also a general characteristic of such quenches related to the fact
that the initial state is not symmetric under space reflections. We
therefore see that the only relevant information about the initial
state, beyond these general features, is the residue of the kinematical
poles, which is much less information than the full functional form
of the overlaps and can even be inferred from general thermodynamic
properties in the two halves of the system.

Whereas the effectively non-interacting character of the system can
be exploited to reduce the problem to a single particle quantum mechanics
problem from which any observable can in principle be derived using
Wick's theorem, we have shown that the mathematical steps used in
the single particle problem can be extended and applied successfully
also in the many particle case. The main point where we have taken
advantage of the non-interacting character of the problem was in the
calculation of the initial state overlaps by means of the Andr\'{e}ief
identity, which undoubtedly is the crucial starting point of the method.
However, as already mentioned in the introduction, for the geometric
quench protocol these overlaps are explicitly known for any value
of the interaction by other means.

On the other hand, in the non-interacting case, summing over the quantised
momentum eigenvalues in order to perform the expansion in the post-quench
basis was done relatively easily as these are explicitly known and
simply equidistant. This is no longer true in the interacting case,
where the quantisation conditions are given by the complicated
set of Bethe equations. However, the complex analysis method for writing
the sum over eigenstates as a contour integral, completely circumvents
the problem of explicit knowledge of the eigenvalues, therefore it
is equally suitable for the analysis of the general case.

These observations are good indications that our method can be generalised
to the genuinely interacting case, which will be the subject of subsequent
work.

\acknowledgements
I would like to thank Alessandro Giuliani for useful discussions. 
This work was supported by the Slovenian Research Agency (ARRS) under grant QTE (N1-0109) 
and by the ERC Advanced Grant OMNES (694544).

\bibliographystyle{SciPost_bibstyle}
\bibliography{QT,bibtexlist}
\selectlanguage{english}%

\end{document}